\documentclass[fleqn,usenatbib]{mnras}

\usepackage{multirow}
\usepackage[T1]{fontenc}

\usepackage{graphicx}
\usepackage{subcaption}
\usepackage{amsmath}
\usepackage{graphicx}
\usepackage{amssymb,txfonts}
\usepackage{xspace}

\newcommand{\nustar}{{\textit{NuSTAR}}\xspace}
\newcommand{\nicer}{{\textit{NICER}}\xspace}
\newcommand{\source}{{MAXI J1820+070}\xspace}
\newcommand{\msun}{{\rm M}_{\sun}}

\def\gtsima{$\; \buildrel > \over \sim \;$}
\def\ltsima{$\; \buildrel < \over \sim \;$}
\def\gsim{\lower.5ex\hbox{\gtsima}}
\def\lsim{\lower.5ex\hbox{\ltsima}}

\title[Stratified hot accretion flow]{A spectrally stratified hot accretion flow in the hard state of MAXI J1820+070}

\author[M. A. Dzie{\l}ak et al.]{
Marta A. Dzie{\l}ak,$^{1}$\thanks{E-mail: mdzielak@camk.edu.pl}
Barbara De Marco$^{2,1}$ and
Andrzej A. Zdziarski$^{1}$
\\
$^{1}$Nicolaus Copernicus Astronomical Center, Polish Academy of Sciences, Bartycka 18, PL-00-716 Warszawa, Poland\\
$^{2}$Departament de F{\'{\i}}sica, EEBE, Universitat Polit{\`e}cnica de Catalunya, Av.\ Eduard Maristany 16, E-08019 Barcelona, Spain
}

\pubyear{2021}

\begin{document}
\label{firstpage}
\pagerange{\pageref{firstpage}--\pageref{lastpage}}
\maketitle

\begin{abstract}
We study the structure of the accretion flow in the hard state of the black-hole X-ray binary MAXI~J1820+070 with \nicer data. The power spectra show broadband variability which can be fit with four Lorentzian components peaking at different time scales. Extracting power spectra as a function of energy enables the energy spectra of these different power spectral components to be reconstructed. We found significant spectral differences among Lorentzians, with the one corresponding to the shortest variability time scales displaying the hardest spectrum. Both the variability spectra and the time-averaged spectrum are well-modelled by a disc blackbody and thermal Comptonization, but the presence of (at least) two Comptonization zones with different temperatures and optical depths is required. The disc blackbody component is highly variable, but only in the variability components peaking at the longest time scales ($\gsim1$ s). The seed photons for the spectrally harder zone come predominantly from the softer Comptonization zone. Our results require the accretion flow in this source to be structured, and cannot be described by a single Comptonization region upscattering disc blackbody photons, and reflection from the disc. 
\end{abstract}
\begin{keywords}
accretion, accretion discs--X-rays: binaries--X-rays: individual (MAXI J1820+070)
\end{keywords}

\section{Introduction}
\label{sec:intro}

The majority of known black-hole (BH) X-ray binaries (XRBs) are transient \citep{Coriat2012MNRAS.424.1991C}. They spend most of their time in a quiescent state, characterized by low/undetectable levels of the X-ray luminosity, $L_{\rm X}/L_{\rm Edd}\lesssim 10^{-5}$ (where $L_{\rm Edd}$ is the Eddington luminosity). After some years of quiescence, they go through X-ray brightening episodes, or outbursts, lasting from a few to tens of months. During an outburst the X-ray spectral and timing properties change dramatically \citep[e.g.][]{Belloni2005A&A...440..207B, Homan2005Ap&SS.300..107H, Dunn2010MNRAS.403...61D, Muoz2011MNRAS.410..679M,Heil2015MNRAS.448.3339H}, and the source goes through a sequence of different accretion states. At the beginning and at the end of an outburst the source is observed in a hard state, spectrally dominated by a hard X-ray thermal Comptonization component \citep[e.g. a review by][]{Done2007A&ARv..15....1D}. During the hard state the source becomes increasingly brighter at almost constant spectral hardness, typically over periods of weeks to a few months. Apart from the dominant hard X-ray Comptonization component, broad band X-ray data show spectral complexity, requiring additional components. Many hard state spectra show a soft component, which can be ascribed, at least partly, to the hardest part of the disc thermal emission  \citep[due to intrinsic dissipation and/or hard X-ray irradiation, e.g.][]{ZdziarskiDeMarco2020ApJ...896L..36Z}, with a typical inner temperature of $kT_{\rm in}\sim 0.1$--0.3 keV. At higher energies, signatures of disc reflection are clearly observed. Fits with relativistic models have yielded controversial results (see \citealt{Bambi20} for a review). Some of these studies support the presence of a disc reaching the immediate vicinity of the innermost stable circular orbit (ISCO) already at relatively low luminosities in the hard state \citep[e.g][]{Miller2006ApJ...653..525M, Miller2008ApJ...679L.113M, Reis2008MNRAS.387.1489R, Reis2010MNRAS.402..836R, Tomsick2008ApJ...680..593T, Petrucci2014A&A...564A..37P, Garcia2015ApJ...813...84G, WangJi2018ApJ...855...61W}. Other studies are instead in agreement with a disc truncated at large radii (a few tens of $R_{\rm g}$, where $R_{\rm g}=GM/c^2$ is the gravitational radius and $M$ is the BH mass) from low to high hard-state luminosities \citep[e.g.][]{ McClintock2001ApJ...555..477M, Esin2001ApJ...555..483E, DoneDiaz2010MNRAS.407.2287D, Kolehmainen2014MNRAS.437..316K, Plant2015A&A...573A.120P, BasakZdziarski2016MNRAS.458.2199B, Dzielak2019MNRAS.485.3845D}. The latter results are in line with theoretical models which predict evaporation of the inner disc \citep{Meyer2000A&A...361..175M, Petrucci2008MNRAS.385L..88P, Begelman2014ApJ...782L..18B, Kylafis2015A&A...574A.133K, Cao2016ApJ...817...71C}. 

One way to understand the origin of these discrepancies is to consider more complexity in the underlying Comptonization continuum. The continuum is usually assumed to result from scattering of disc photons in an homogeneous region near the BH, which in most cases results in a flat or convex spectral shape. On the other hand, inhomogeneity of the Comptonization region is able to produce additional curvature, resulting in a concave underlying continuum, as often required by the data \citep{DiSalvo2001ApJ...547.1024D, Frontera2001ApJ...546.1027F, Ibragimov2005MNRAS.362.1435I, Makishima2008PASJ...60..585M, Nowak2011ApJ...728...13N, Yamada2013PASJ...65...80Y, Basak2017MNRAS.472.4220B, Zdziarski21}. Such conditions can account for both a part of the Fe K line red wing and a part of the reflection hump at $\gtrsim$10\,keV, which allow the need for extreme relativistic reflection parameters, as well as a super-solar Fe abundance \citep[often found in the fits,][]{Furst2015ApJ...808..122F,Garcia2015ApJ...813...84G, Parker2015ApJ...808....9P, Parker2016ApJ...821L...6P, Walton2016ApJ...826...87W, Walton2017ApJ...839..110W,  Tomsick2018ApJ...855....3T} to be relaxed. 
The presence of a radially stratified and spectrally inhomogeneous Comptonization region is independently supported by X-ray variability studies. In particular, the common detection of frequency-dependent hard X-ray lags (hard X-ray variations lagging soft X-ray variations) in energy bands dominated by the primary hard X-ray continuum can be explained by mass accretion rate fluctuations propagating inward \citep{Lyubarskii1997MNRAS.292..679L, Arevalo2006MNRAS.367..801A}. However, in order to detect a net hard lag, the zone these perturbations propagate through must have a radially-dependent emissivity profile, with the inner regions emitting a harder spectrum \citep{Kotov2001MNRAS.327..799K}. Spectral-timing techniques, which allow resolving the spectral components contributing to variability on different timescales (frequency-resolved spectroscopy) have been applied to BH XRBs, showing that those spectra generally harden with decreasing time scale. This also provides evidence for the differences in the observed spectral shape to be related to their different distances from the BH, see, e.g.   \citet{Revnivtsev1999A&A...347L..23R, Axelsson2018MNRAS.480..751A}. Following these findings, \citet{Mahmoud2018MNRAS.480.4040M} and \citet{Mahmoud2019MNRAS.486.2137M} created theoretical models reproducing those observations by multi-zone models coupled with propagation mass accretion rate fluctuations.

We study the BH XRB system \source, focusing on two {\it Neutron star Interior Composition ExploreR} (\nicer; \citealt{Gendreau16}) observations taken during the rise of its 2018 outburst \citep[see][for a systematic analysis of all the first part of the outburst]{DeMarco21}. We perform frequency-resolved spectroscopy method in the formulation outlined by \citet{Axelsson2018MNRAS.480..751A}. We fit frequency-resolved energy spectra using Comptonization models, with the goal to constrain the structure of the hot flow, and verify the possible presence of multiple Comptonization zones.

\source was first detected in the optical wavelengths by the All-Sky Automated Search for SuperNovae on 2018-03-06 \citep[ASASSN-18ey,][]{Tucker2018ApJ...867L...9T}. It was detected in X-rays five days later by the Monitor of All-sky X-ray Image (MAXI; \citealt{Matsuoka09}) on board of {\it International Space Station} \citep{Kawamuro2018ATel11399....1K}. The two detections were connected to the same source by \citet{Denisenko2018ATel11400....1D}. It was then identified as a BH candidate \citep{Kawamuro2018ATel11399....1K,Shidatsu2019ApJ...874..183S}. \citet{Torres2019ApJ...882L..21T} confirmed the presence of a BH by dynamical studies of the system in the optical band, and \citet{Torres2020ApJ...893L..37T} measured in detail the parameters of the system. The donor is a low-mass K star and the orbital period is $0.68549 \pm 0.00001$\,d \citep{Torres2019ApJ...882L..21T}.

The BH mass is $M\approx (5.95\pm 0.22)\msun/\sin^3 i$ \citep{Torres2020ApJ...893L..37T}, anticorrelated with the binary inclination, $i$. Those authors estimated the latter as $66\degr<i<81\degr$. On the other hand, \citet{Atri2020MNRAS.493L..81A} estimated the inclination of the jet\footnote{We note that the two inclinations can be different, and, in fact, are significantly different in some BH XRBs, e.g. GRO 1655--40, \citep[][]{Hjellming95, Beer02}.} as $i\approx 63\pm 3\degr$. Based on their radio parallax, \citet{Atri2020MNRAS.493L..81A} determined the distance to \source as $d\approx 3.0\pm 0.3$\,kpc, consistent with the Gaia Data Release 2 parallax, yielding $d\approx 3.5^{+2.2}_{-1.0}$\,kpc \citep{Bailer18,Gandhi19, Atri2020MNRAS.493L..81A}. \source is a relatively bright source, with the peak 1--100\,keV flux estimated by \citet{Shidatsu2019ApJ...874..183S} as $\approx\! 1.4\times 10^{-7}$\,erg\,cm$^{-2}$\,s$^{-1}$, which corresponds to the isotropic luminosity of $\approx\! 1.5(d/3\,{\rm kpc})^2\times 10^{38}$\,erg\,s$^{-1}$. Assuming an inclination $i=66\degr$ this corresponds to $\sim$15\%  $L/L_{\rm{Edd}}$.
The value of Galactic hydrogen column density toward the source has been estimated in the range of (0.5--$2)\times 10^{21}$\,cm$^{-2}$ \citep{Uttley2018ATel11423....1U, Kajava2019MNRAS.488L..18K, Fabian2020MNRAS.493.5389F, Xu2020ApJ...893...42X}.

\section{Data reduction}
\label{sec:data_red}

For our analysis, we choose two relatively long observations of \source carried out by \nicer, see Table~\ref{tab:obsid}. Hereafter, we refer to the studied observations using only the last three digits of their identification number (ID), i.e. 103 and 104.
They correspond to the initial phases of the 2018 outburst, when the source was in the rising hard state (see Figure 1 in \citealt{DeMarco21}). We estimate\footnote{We have used a model including a disc blackbody plus a power law, both absorbed by the cold gas in the interstellar medium; \texttt{TBabs(diskbb+powerlaw)} in {\sc Xspec}.} their energy fluxes to be $\approx$6.5 and $\approx9.5 \times 10^{-9}$\,erg\,cm$^{-2}$\,s$^{-1}$, respectively, in the 0.3--10\,keV energy band.

The data were reduced using the {\tt NICERDAS} tools in {\sc heasoft} v.6.25, starting from the unfiltered, calibrated, all Measurement/Power Unit (MPU) merged files, {\tt ufa}. We applied the standard screening criteria \citep[e.g.][]{Stevens2018} through the {\tt NIMAKETIME} and {\tt NICERCLEAN} tools. We check for periods of high particle background, i.e., with rate $>$2\,count\,s$^{-1}$, by inspecting light curves extracted in the energy range of 13--15\,keV, in which the contribution from the source is negligible because of the drop in the effective area \citep{Ludlam2018}.

Of the 56 focal plane modules (FPMs) of the \nicer X-ray Timing Instrument (XTI), four (FPMs 11, 20, 22, and 60) are not operational. Additionally, we filter out the FPMs 14 and 34, since they are found to occasionally display increased detector noise\footnote{\url{https://heasarc.gsfc.nasa.gov/docs/nicer/data\_analysis/nicer\_analysis\_tips.html}}. Thus, the number of FPMs used for our analysis is 50. In order to correct for the corresponding reduction of effective area, the reported source fluxes were rescaled by the number of used FPMs.

The shortest good time intervals, with the length $<10$\,s, were removed from the analysis, resulting in the net, on-source, exposure times reported in Table~\ref{tab:obsid}. Filtered event lists were barycentre-corrected and used to extract the light curves and spectra using {\tt XSELECT} v.2.4e. We used publicly distributed ancillary response and redistribution matrix files\footnote{\url{https://heasarc.gsfc.nasa.gov/docs/heasarc/caldb/data/nicer/xti/index.html}} as of 2020-02-12 and 2018-04-04, respectively. Fits were performed using the X-Ray Spectral Fitting Package ({\sc Xspec} v.12.10.1; \citealt{Arnaud96}). Hereafter uncertainties are reported at the 90 per cent confidence level for a single parameter.

\begin{table}
\centering
\caption{The log of the \nicer observations used for the analysis. Exposures correspond to the effective on-source time after data cleaning.}
\label{tab:obsid}
\begin{tabular}{ c c c }
 \hline
 Obs.\ ID & Start time & Exposure  \\ 
 	     & [yyyy-mm-dd hh:mm:ss] & [s]   \\
 \hline
 1200120\textbf{103} & 2018-03-13 23:56:12 & 9474  \\
 1200120\textbf{104} & 2018-03-15 00:36:04 & 6231  \\
\hline
\end{tabular}
\end{table}

\section{Analysis}
\subsection{Power spectral density}
\label{sec:timing analysis}

We first compute and fit the power spectral density (PSD) for each of the observations in two broad energy bands, 0.3--2\,keV and 2--10\,keV, in order to identify components contributing to the observed variability on different timescales. We extract light curves with a time bin of 0.4\,ms in these energy bands. The light curves were split into segments of 200\,s length. We calculate the PSD of each segment and average them, in order to obtain more accurate estimates of the PSD of each single observations. The chosen time bin and the segment length allow us to cover the range of frequencies of 0.005--1250\,Hz. This range well samples both the broad-band noise intrinsic to the source and the Poisson noise level. The latter was fit at frequencies of $\nu>300$\,Hz and subtracted from the PSDs at all frequencies. We adopt a logarithmic rebinning in order to improve the statistics at high frequencies.

\begin{figure}
 \begin{subfigure}[b]{0.495\textwidth}
	\includegraphics[width=0.94\columnwidth]{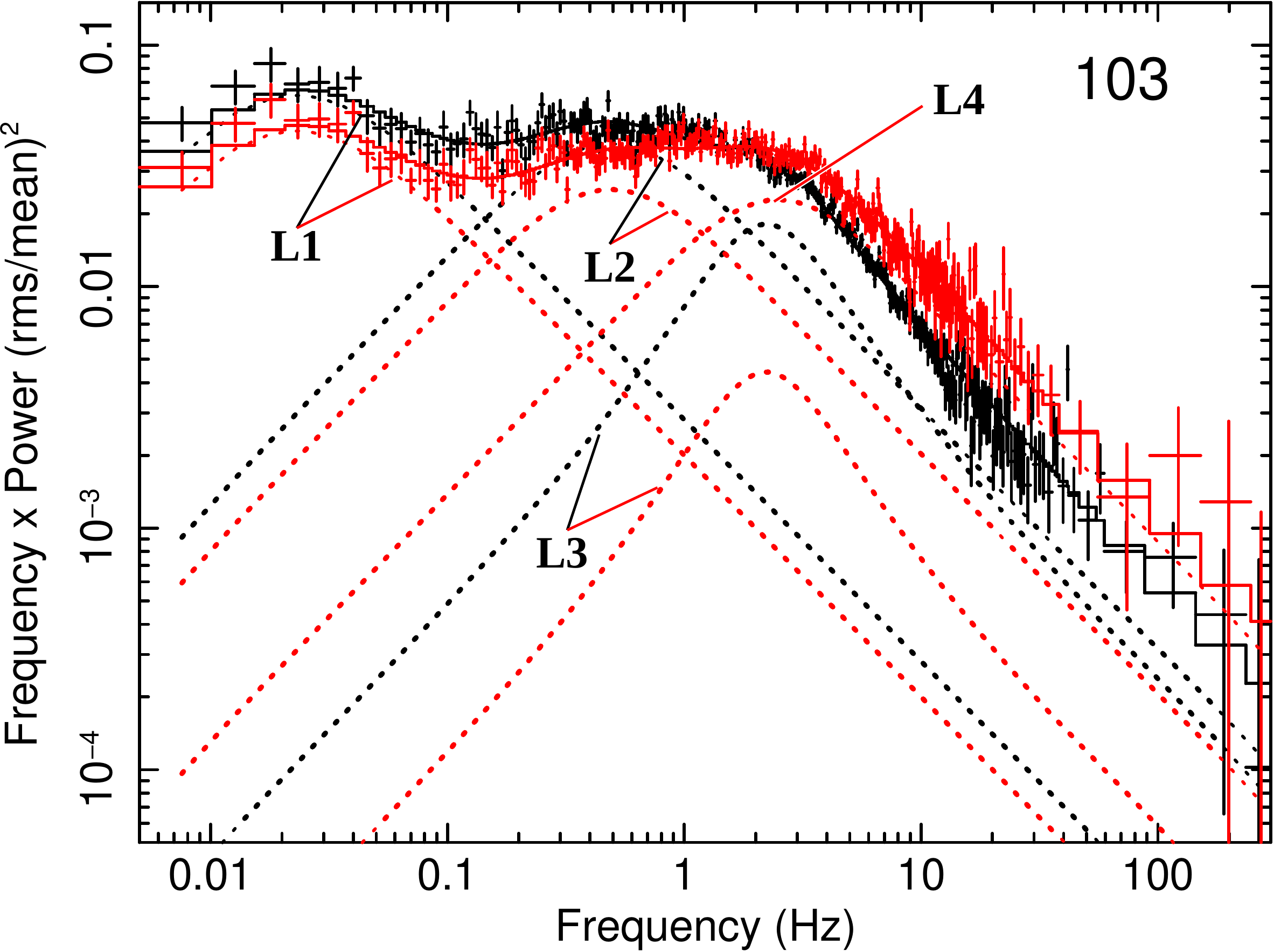}
		\end{subfigure}
	%\vspace{1.pt}
 \begin{subfigure}[b]{0.495\textwidth}
	\includegraphics[width=0.94\columnwidth]{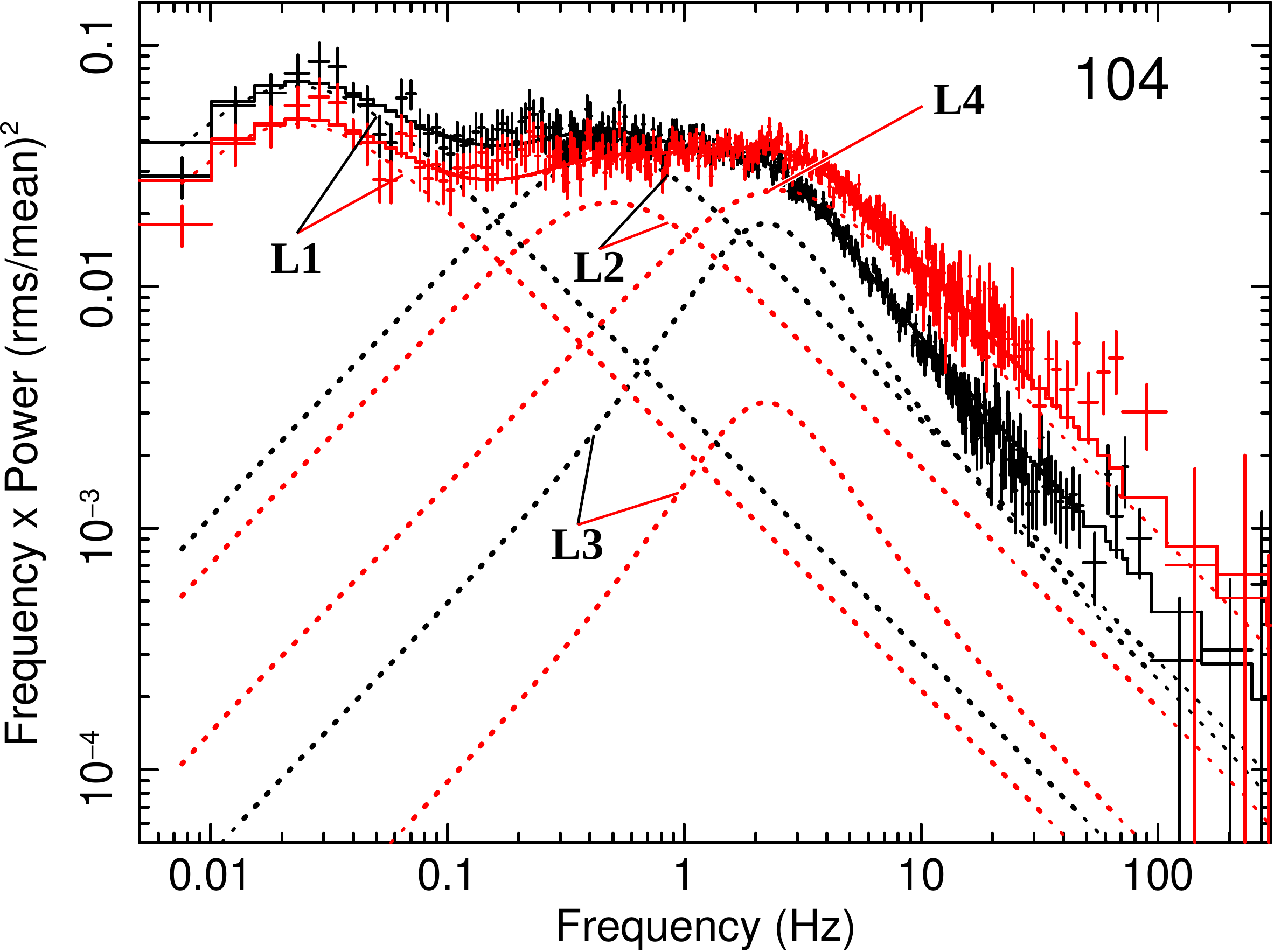}
		\end{subfigure}
    \caption{The Poisson noise subtracted PSDs of the two analyzed observations in the two energy bands, 0.3--2\,keV (black symbols) and 2--10\,keV (red symbols). The total best-fit models (solid lines) and the Lorentzian components of the model (dotted curves) are also shown.}
\label{fig:pspectra}
\end{figure}

We normalize the PSDs using the fractional squared root-mean-square units \citep[(\emph{rms}/mean rate)$^2$Hz$^{-1}$, ][]{Belloni1990,Miyamoto1992}. The PSDs are shown in Fig.\ \ref{fig:pspectra}. Their complex structure hints at the presence of several variability components. As commonly done in the literature \citep{Belloni1997A&A...322..857B, Nowak2000MNRAS.318..361N}, we fit each PSD with a sum of Lorentzian components, for which we use {\sc Xspec}. A Lorentzian is described by
\begin{equation}
    P(\nu) = K\frac{\sigma }{2\pi }\frac{1}{(\nu - \nu_{0})^{2}+ (\sigma/2)^{2}},
\label{Lorentz}
\end{equation}
where $K$ is the normalization, $\sigma$ is the full width at half maximum, and $\nu_{0}$ is the centroid frequency. However, the maximum power (i.e. the peak of $\nu P(\nu)$) is observed at the frequency $\nu_{max}>\nu_0$ \citep{Belloni1997A&A...322..857B}
\begin{equation}\label{numax}
    \nu _{\rm{max}} = \sqrt{\nu _{0}^{2} + \left(\frac{\sigma}{2}\right)^{2} }.
\end{equation}

We jointly fit the resulting four PSDs, we tie $\nu_0$ and $\sigma$ parameters, but allow for different normalization. We start with fitting a single Lorentzian, and add a new one if significant residuals remained. We find that the hard band is well described by four Lorentzian components, but an addition of the highest-frequency, fourth, Lorentzian in the soft band PSD yields a normalization consistent with zero. We also find that letting $\nu_0$ and $\sigma$ vary independently for each PSD does not lead to significant fit improvements. Therefore, our final joint model consists of four Lorentzians (hereafter L1, L2, L3, and L4), with the parameters given in Table \ref{tab:lore}. The Poisson noise subtracted PSDs with the best-fit models are shown in Fig.\ \ref{fig:pspectra}.

We check that the PSDs of the two observations have compatible shape and consistent fractional normalization in each of the two energy bands ( $\approx$32 and $\approx$27 per cent in the soft and hard band, respectively). This means that there is no significant deviation from stationarity between the two observations. Therefore, we decide to combine the two observations in order to obtain a higher signal-to-noise for the remainder of our analysis.

\renewcommand{\arraystretch}{1.2}
\begin{table}
\centering
\caption{The results of the joint fits of the PSDs of the observations 103 and 104 in the 0.3--2\,keV and 2--10\,keV energy bands with the model consisting of four Lorentzians, see equation (\ref{Lorentz}). We also give the resulting peak frequency of $\nu P(\nu)$, equation (\ref{numax}), and the resulting reduced $\chi^2$.
}
\label{tab:lore}
\begin{tabular}{ccccc}
\hline
        & L1   & L2   & L3   & L4  \\
\hline
$\nu_{\rm{0}}$ [Hz] &  0$^{+0.006}$   & 0.095$^{+0.038}_{-0.039}$    & 1.59$^{+0.16}_{-0.18}$    &  0.65$^{+0.27}_{-0.65}$  \\
$\sigma$ [Hz]   & 0.045$^{+0.004}_{-0.003}$    & 0.96$^{+0.09}_{-0.08}$    & 3.07$^{+0.11}_{-0.09}$    & 4.80$^{+1.01}_{-0.35}$   \\
$\nu_{\rm{max}}$ [Hz] &  0.023$^{+0.002}_{-0.002}$   & 0.489$^{+0.045}_{-0.039}$    & 2.21$^{+0.12}_{-0.13}$    &  2.49$^{+0.49}_{-0.24}$  \\ 
$\chi^{2}_{\nu}$     & \multicolumn{4}{c}{0.993} \\
\hline
\end{tabular}
\end{table}

\subsection{Extraction of frequency-resolved spectra}
\label{sec:spec-tim analysis}
\begin{figure*}

	 \begin{subfigure}[b]{0.495\textwidth}
	\includegraphics[width=0.95\columnwidth]{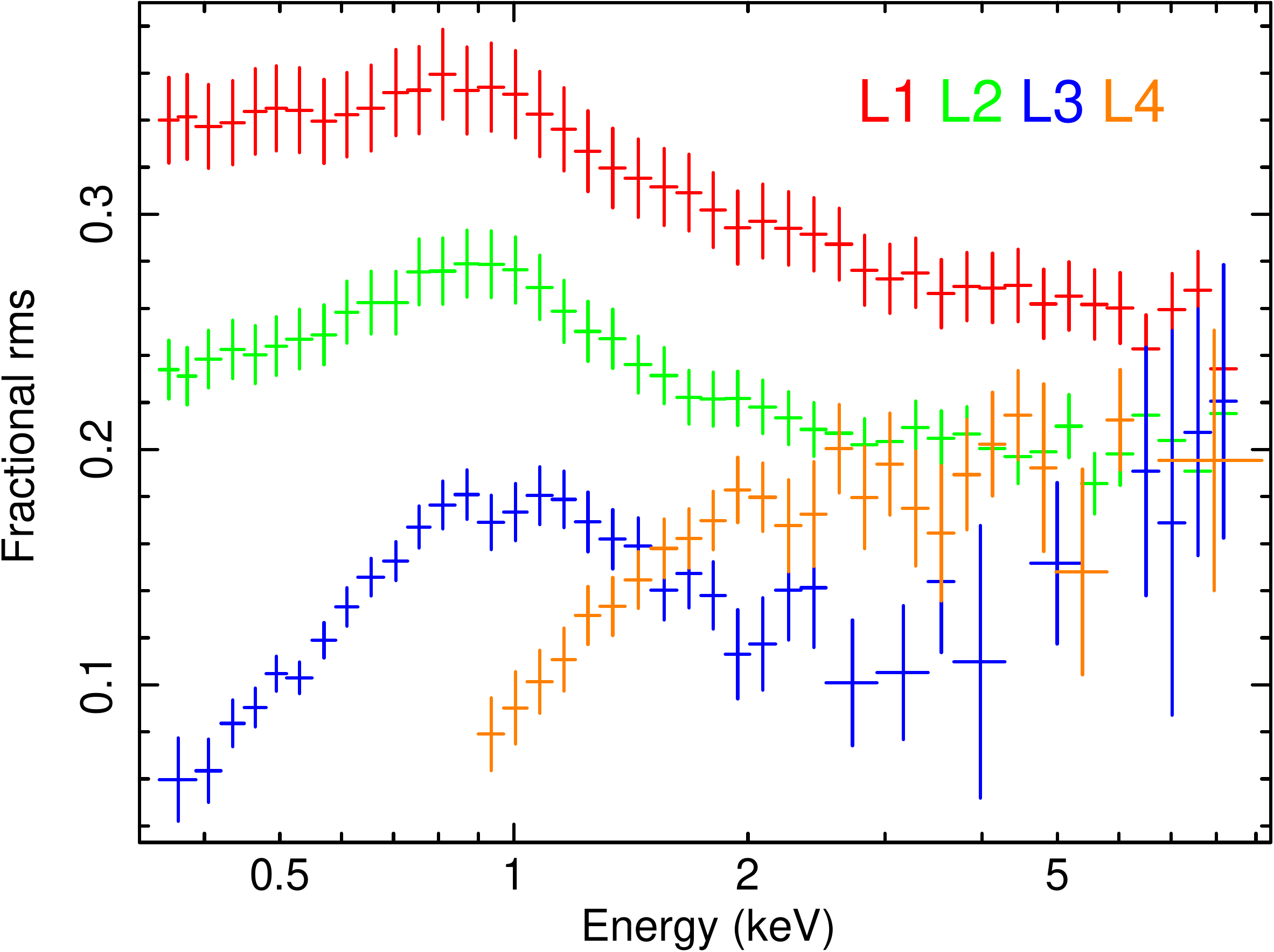}
		\end{subfigure}
 \begin{subfigure}[b]{0.495\textwidth}
	\includegraphics[width=0.95\columnwidth]{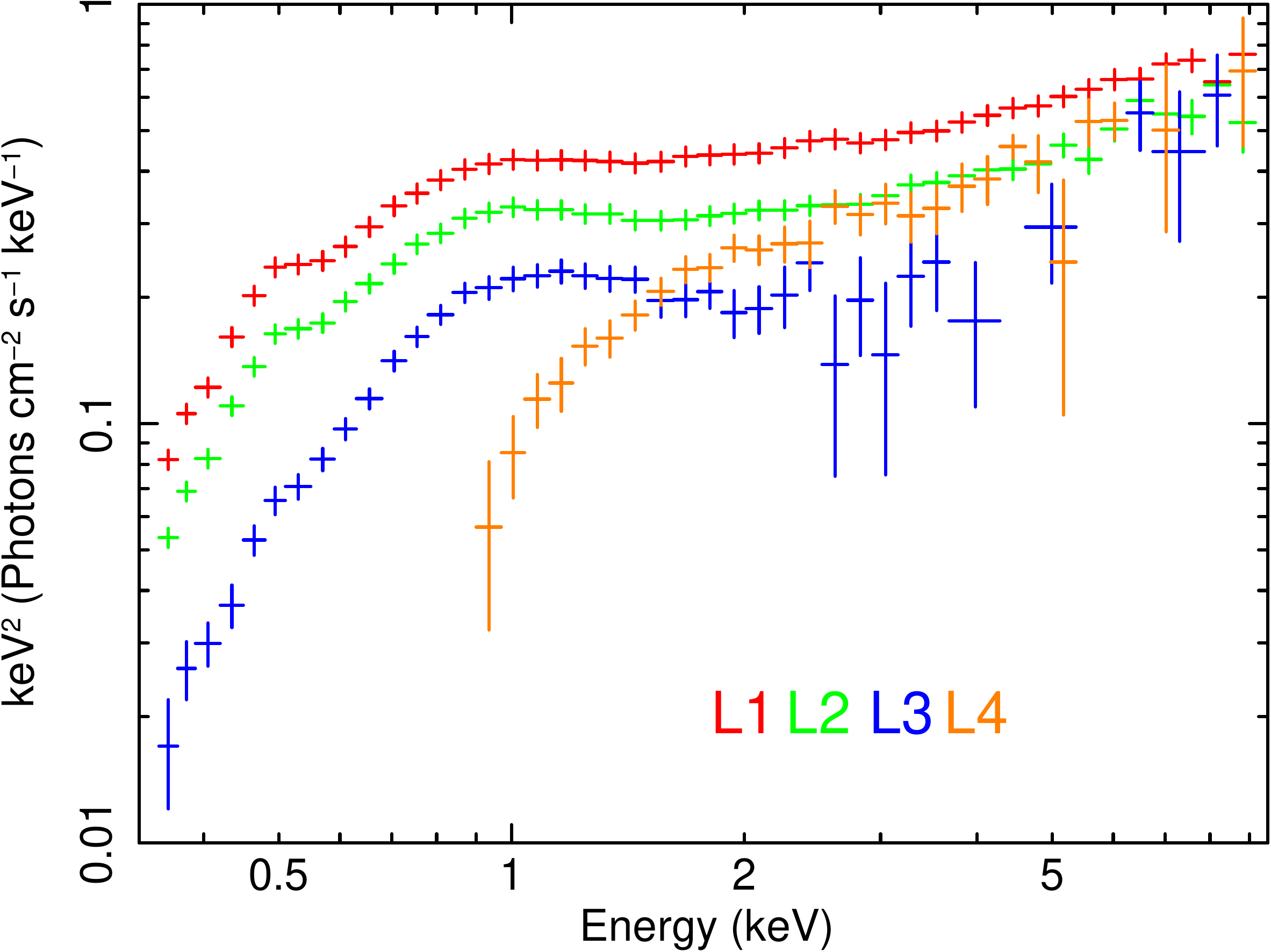}
		\end{subfigure}
    \caption{The fractional \emph{rms} spectra (left panel) and absolute \emph{rms} spectra (right panel) of each of the four Lorentzian components, unfolded to a constant model.}
\label{fig:erms}
\end{figure*}

We extract the fractional \emph{rms} and absolute \emph{rms} energy spectra using the Lorentzian fits presented in Section \ref{sec:timing analysis}. The fractional \emph{rms} spectrum shows the distribution of fractional variability power as a function of energy \citep[e.g.][]{Vaughan2003MNRAS.345.1271V} and can be used to readily assess the presence of spectral variability (e.g. presence of multiple variable spectral components, spectral pivoting). On the other hand, the absolute \emph{rms} spectrum shows the spectral shape of the components that contribute to variability, and can be used to directly fit spectral models. However, this comes with some caveats that will be discussed in Sect.~\ref{sec:discussion}.
We use the technique proposed by \cite{Axelsson2013MNRAS.431.1987A}. We create a logarithmic energy grid with 47 bins in the 0.3--9.9\,keV energy range. We then calculate the PSD in fractional (\emph{rms}/mean rate)$^2$Hz$^{-1}$ and absolute units \citep[({count~s}$^{-1}$)$^2$Hz$^{-1}$,][]{Vaughan2003MNRAS.345.1271V} for each of the bins. We then fit the PSD in each band with the same Lorentzian components as the best-fit model to the PSD of the broad energy bands (Table \ref{tab:lore}), keeping the values of $\nu_0$ and $\sigma$ fixed, and allowing only the normalization of each Lorentzian to vary. The best-fit PSD models for each narrow energy band are then used to extract the fractional and absolute \emph{rms} variability amplitude spectrum of each Lorentzian. To this aim we analytically integrate the best-fit Lorentzians within the frequency range 0.001--1000\,Hz, such that:
\begin{equation}
rms=\sqrt{\left(\frac{K}{\pi}\right)\arctan\left(\frac{(\nu-\nu_0)}{\sigma/2}\right)\bigg\rvert_{\nu = 0.001}^{\nu=1000}}.    
\end{equation}
We plot these values as a function of energy to obtain the fractional and absolute \emph{rms} spectra. Results are reported in Fig.\ \ref{fig:erms}. We note that since these four components were obtained from the power spectrum, which gives \emph{rms}$^2$, they were assumed to be completely uncorrelated, and, in order to obtain the total \emph{rms} spectrum, the single \emph{rms} spectra have to be summed in quadrature. We note that the \emph{rms} spectra have the redistribution matrix obtained by rebinning the matrix for the average spectrum.

The errors on the \emph{rms} spectra were calculated based on the $1\sigma$ deviation of the normalization of the best-fit Lorentzians. To this aim, for each Lorentzian and in each energy band, we draw two random numbers distributed according to a Gaussian distribution with the mean equal to the best-fit value of the normalization, $K$, and the standard deviation equal to its $1\sigma$ lower/upper error of $K$. This procedure was repeated 100 times for each Lorentzian component, resulting in two sets of 100 simulations each. We use each set to estimate the upper and lower limits on the \emph{rms} in every energy channel. In order to have symmetric errors (as required in {\sc Xspec}) we take the average of the obtained upper and lower errors. The unconstrained points, i.e., those resulting in upper limits on the \emph{rms} (e.g. as in the case of the normalization of L4 in the softest energy bands) were omitted in the fits and are not presented in the plots. 

The fractional \emph{rms} spectra of each Lorentzian show significant spectral variability (Fig.~\ref{fig:erms}, left). In particular, in the low frequency components (L1, L2, and L3) a local peak of variability is observed at around 1 keV (at the peak the fractional \emph{rms} decreases from $\sim$36\% to $\sim$18\% from low to high frequencies). Then, on either side of the peak the fractional rms tends to decrease. While for L1 and L2 the decrease is steady up to the highest sampled energies, for L3 a reversal of this trend is observed above $\sim$3 keV. L4 does not show any low energy peak, and the variability at around $\sim$1 keV results suppressed. 
In L1, L2, and L4 the fractional variability appears to flatten out at energies $\gtrsim$3 keV (to a level of $\sim$26\% for L1, and $\sim$20\% for L2 and L4). This suggests that high energy spectral components predominantly vary in normalization only. 
We note that these results are in agreement with those reported by \cite{Axelsson_Veledina2021arXiv210308795A}.

\subsection{Spectral fits of absolute \emph{rms} and time-averaged spectra}
\label{sec:fits}
We then proceed to fit absolute \emph{rms} and time-averaged spectra, using standard {\sc Xspec} models. In our models, absorption due to the interstellar medium (ISM) is modeled using \texttt{tbabs} \citep{Wilms2000ApJ...542..914W}, with the elemental abundances from the same paper. We model thermal Comptonization using the model \texttt{ThComp} \citep{Z20_thcomp}. Since this is a convolution model, the covered energy range has to be extended beyond that of the \nicer spectra in order to get a full range of energies for the seed photons; we use\footnote{The {\sc Xspec} commands \texttt{energies extend low 0.001} and \texttt{energies extend high 2000} were used. This avoids modifying the energy grid of the data.} the range of 0.01\,keV--2000\,keV. Since at high energies the sensitivity of the instrument is limited to $E\lsim10$ keV, we are not able to fit the electron temperature, and we fix it at $kT_{\rm{e}}=50$ keV, which closely resembles the values found from simultaneous fits of \nustar data \citep{Zdziarski21b}. In our modelization, we assume that all the parameters governing the shape of the spectral components are constant and that the variability associated within each Lorentzian is predominantly due to variations in normalization of the spectral components. In particular, our modelization discards spectral pivoting, as suggested by fractional \emph{rms} spectra (Fig.~\ref{fig:erms}, left; see also Discussion in Sect.~\ref{sec:discussion}).

\begin{figure} \begin{subfigure}[b]{0.45\textwidth} 
\includegraphics[width=0.96\columnwidth]{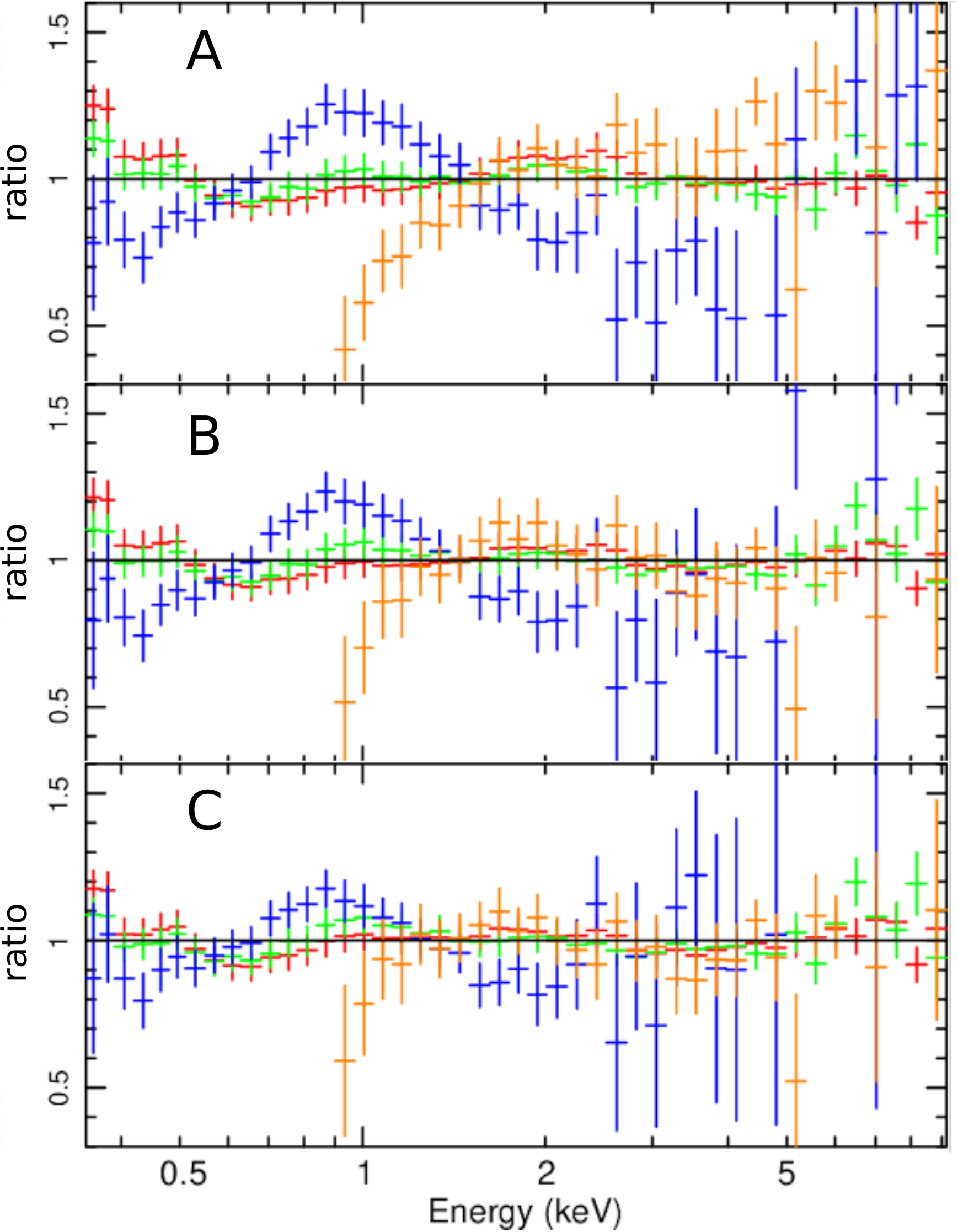} \end{subfigure} \caption{The 
data-to-model ratios of models A, B, C (from top to bottom; see Table~\ref{tab:onec_model}). The same colour coding as in Fig.\ \ref{fig:erms} is used: L1 -- red, L2 -- green, L3 -- blue and L4 -- orange. } \label{fig:resi} \end{figure}

We first consider single Comptonization models, with the disc thermal emission \citep{Mitsuda84} being the source of seed photons. The ISM hydrogen column density is fixed at $N_{\rm H}=1.4\times 10^{21}{\rm cm^{-2}}$ (which is the best-fit value obtained from our fit to the time-averaged spectrum, see Table \ref{tab:twoc_model} below). {\tt ThComp} has the fraction of the seed photons incident on the Comptonization region, $f$, as a free parameter. However, we hereafter set $f=1$ and include a separate disc blackbody component with the same parameters in order to separate the two components in the plots and tables. The {\sc Xspec} notation of the model is \texttt{TBabs(diskbb$_1$+ThComp(diskbb$_2$))}. We note that such additive form of the variability model spectrum assumes that the two spectral components are fully correlated. In our first case, Model A, we fit \emph{rms} spectra of each Lorentzian (L1, L2, L3, L4) with the same parameters except for the normalizations. The normalization of {\tt thComp} is given by the normalization of the convolved {\tt diskbb} model. The other free parameters are the inner disc temperature, $kT_{\rm in}$, and the low-energy spectral index of the Comptonization spectrum, $\Gamma$ (defined by the energy flux, $F(E)\propto E^{1-\Gamma}$). The obtained best-fit parameters are given in Table \ref{tab:onec_model} and the ratios of the data to the model are shown in Fig.\ \ref{fig:resi} (top panel). We find that this model provides a rather poor description of the data, $\chi_\nu^{2}\approx 318.8/160$. 

We thus check if a more complex spectral structure can improve the fits. In Model B, we allow for the photon index, $\Gamma$, to vary among different Lorentzians. This means that we assume the presence of four different Comptonization zones with different Thomson optical depths, $\tau_{\rm T}$, since the spectral index is a function of both $kT_{\rm e}$ and $\tau_{\rm T}$ (see eq.\ 14 in \citealt{Z20_thcomp}; and \citealt{ST80} for a general discussion). 

In Model C, we also allow for changes among $kT_{\rm{in}}$ of the seeds for Comptonization, and allow them to be different from the directly observed disc blackbody. However, we find the seed photons temperature of the \emph{rms} spectrum of L1 to be unconstrained, so we keep it equal to $kT_{\rm{in}}$ of the directly observed disc. Model C assumes the presence of four different disc blackbodies. While it is unrealistic, we include it as a phenomenological description in order to highlight that the data do require the seed photons to have different characteristic energies for different components. The best-fit parameters of Models B and C are given in Table \ref{tab:onec_model}, while the ratios of the data to the corresponding best-fit models are shown in Fig.\ \ref{fig:resi} (middle and bottom panel).

Comparing results from the spectral fits of Model A with those of B and C, we see significant evidence for changes of the spectral properties between the \emph{rms} spectra of the different Lorentzians. In particular, the fit improves when letting an increasing number of parameters free to vary among the different variability components (from model A to model C, $\Delta\chi^{2}\approx -133$). The data show that the two lowest frequency Lorentzians L1 and L2 have very similar spectra, L3 has the softest spectrum (we note that in this case, the poor signal-to-noise at high energies may be the reason for such softening), and L4 has the hardest spectrum. We also observe that the \emph{rms} spectrum of the highest frequency Lorentzian (L4) is best fit by a much higher seed photons temperature ($kT_{\rm{in2}}$ in Table~\ref{tab:onec_model}) than the lower frequency components. This hints at a different source of seed photons for L4, thus suggesting that the innermost parts of the accretion flow are fueled by photons with higher temperature than those fueling the outer parts.

Given the clear indications of changes in spectral properties among different Lorentzian components, we model now the \emph{rms} spectra and the time-averaged spectrum simultaneously, with the aim of obtaining more robust and self-consistent constraints on the structure of the Comptonization region. For consistency with the \emph{rms} spectra, the time-averaged spectra of observations 103 and 104 were averaged (using {\tt MATHPHA} in {\sc ftools}). The resulting spectrum was rebinned so as to have a minimum of 3 original energy channels and signal-to-noise $\geq$50 in each new energy channel. To account for uncertainties in the current calibration, a systematic error of 1 per cent was added. (Note that this step is not required for the \emph{rms} spectra, owing to their lower resolution.) Given the limited bandpass of \nicer, we did not include complex reflection models in the fits, which would result in overfitting the \emph{rms} spectra. Therefore, we model the fluorescent Fe K line with a simple Gaussian component.

\begin{figure}
 \begin{subfigure}[b]{0.46\textwidth}
	\includegraphics[width=1\columnwidth]{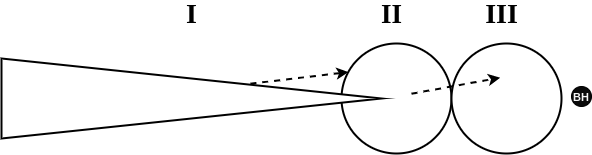}
\end{subfigure}

\vspace{8pt}

 \begin{subfigure}[b]{0.45\textwidth}
	\includegraphics[width=1\columnwidth]{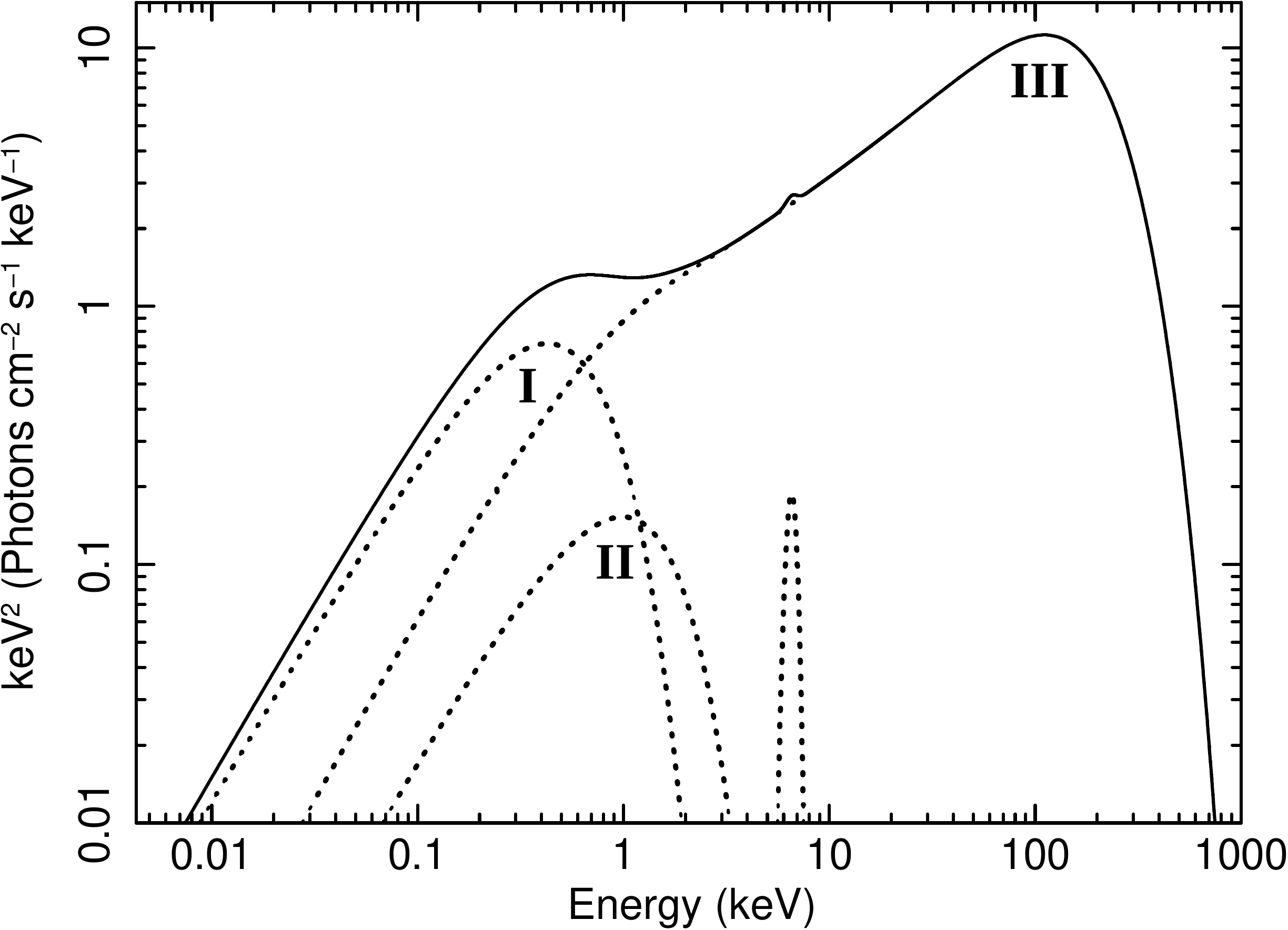}
\end{subfigure}
    \caption{(Top panel): a drawing of the geometry assumed in model D. (Bottom panel) the unabsorbed best-fit to the time-averaged spectrum (solid curve) and its components (dotted curves) marked by the zones shown in the top panel and Table\ \ref{tab:twoc_model}. 
    }
\label{geo}
\end{figure}

\begin{figure}
 \begin{subfigure}[b]{0.43\textwidth}
	\includegraphics[width=1\columnwidth]{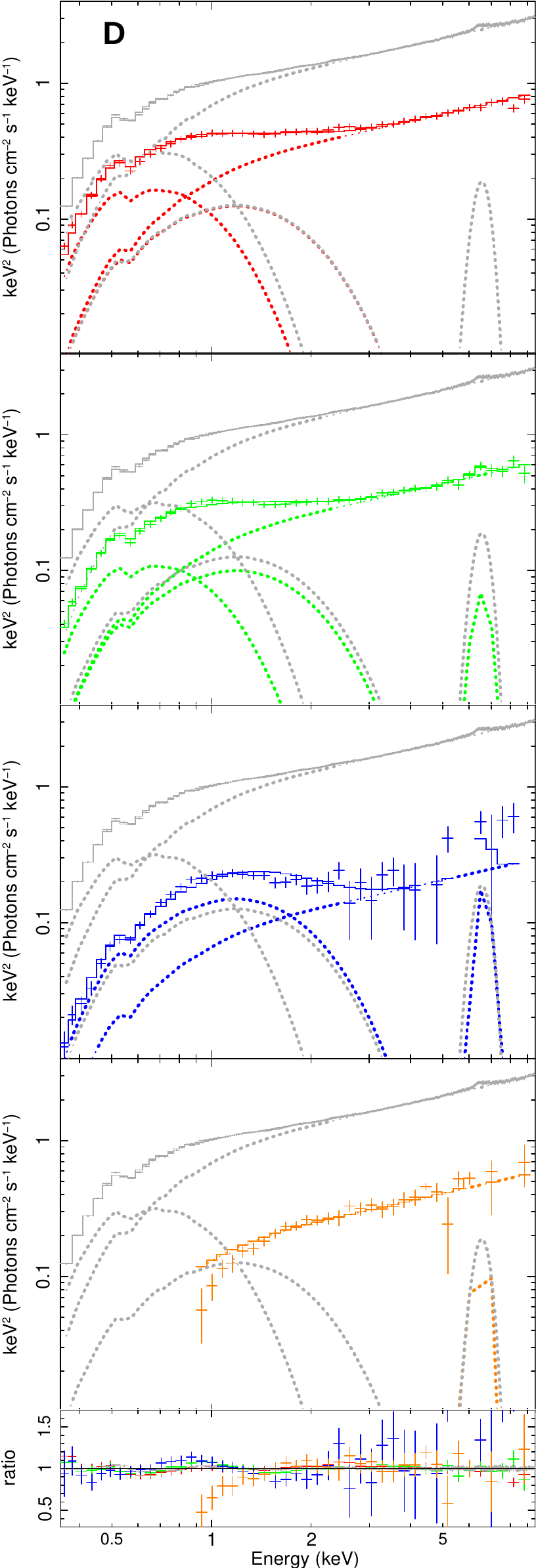}
\end{subfigure}
\caption{The best-fit models and data of the Lorentzian spectral components (from top to bottom: L1, L2, L3, L4), and the data-to-model ratios (bottom panel) of model D. The colour code is the same as in Fig.~\ref{fig:erms}, and the time-averaged spectrum and its residuals are shown in grey.}
\label{fig:d}
\end{figure}

Results obtained from our fits with model C show that the overall spectrum cannot be fit by a single Comptonization component. We therefore test for a more complex model, assuming that the hot flow can be described by two Comptonization zones. The model, denoted as D, comprises a directly visible disc (zone I), which photons are upscattered in the outer Comptonization zone II, and those upscattered photons are partly directly observed and partly are upscattered in the inner Comptonization zone III, as illustrated in the drawing in Fig.\ \ref{geo} (top panel). The {\sc Xspec} notation is given in the caption to Table\ \ref{tab:twoc_model}. Again, the underlying assumption is that the three components are fully correlated. The unabsorbed best-fit model to the time-averaged spectrum with these components is presented in Fig.\ \ref{geo} (bottom panel). Since the Fe line is unconstrained in the \emph{rms} spectra, we require their normalizations to be $\leq$ than that in the time-averaged spectrum. The best-fit parameters are given in Table\ \ref{tab:twoc_model}, and the best-fit model and the data-to-model ratios are shown in Fig.\ \ref{fig:d}. We find that model D can simultaneously describe the time-averaged and \emph{rms} spectra well, with $\chi_\nu^2\approx 370.4/415$. These results can be summarized as follows.

{\bf Low frequency Lorentzians (L1 and L2)}:
In model D, the \emph{rms} spectra of L1 and L2 show significant contributions from all of their spectral components (see normalizations in Table \ref{tab:twoc_model}). Still, most of the photons related to the variability component piking at frequencies $\lesssim$1\,Hz are associated with the direct disc emission, which variable components comprise $\approx$63 per cent of the total (time-average) disc blackbody flux at the best fit. In zone II, the variable components actually exceed the corresponding time-average, which implies an inaccuracy resulting from the assumed spectral model (see Section \ref{sec:discussion}). Still, it indicates that most of the photons in this zone are variable.

{\bf High frequency Lorentzians (L3 and L4)}:
These variability components show no direct emission from the disc at the best fit. This means that the disc gives low contribution to the variability components peaking at the time scales $\gtrsim$1\,Hz. Almost all blackbody photons are instead used as seeds for the Comptonization regions. For L3, $\approx$3/4 of those seed photons are Compton upscattered in the outer Comptonization region (zone II), and $\approx$1/4 of them, in the inner Comptonization region (zone III). On the other hand, for L4, most of the disc photons upscattered in the outer Comptonization region (zone II) are used as seed photons for the inner Comptonization region (zone III).

In the model, emission from the inner Comptonization region dominates the bolometric X-ray flux. The outer Comptonization region is significantly softer than the inner one, i.e., $\Gamma_1>\Gamma_2$. Overall, these results show that the structure for the accretion flow in the hard state is quite complex, with stratified Comptonization regions. 

\setlength{\tabcolsep}{2.9pt}
\begin{table*}
\centering
\caption{The joint fit results for models with a single Comptonization zone for each of the Lorentzians (L1, L2, L3, L4). The {\sc Xspec} notation is \texttt{TBabs(diskbb$_1$+ThComp(diskbb$_2$))}. The normalization of {\tt thComp} is given by the normalization of the convolved {\tt diskbb} model. In model A, the Comptonization plasmas have the same parameters for all of the Lorentzians, and in model B, they can differ in the optical depth, resulting in different spectral indices, $\Gamma$. Model C is a phenomenological one, in which the inner disc temperatures, $kT_{\rm in2}$, of the seed photons for Comptonization can also differ between the Lorentzians. Hereafter, the parameters assumed to be fixed are denoted by (F).
}
\begin{tabular}{c c c c c c c c c c c c c c c c }
 \hline
 &    &\multicolumn{4}{c}{A} &   & \multicolumn{4}{c}{B} & & \multicolumn{4}{c}{C}  \\   
\cline{3-6} \cline{8-11} \cline{13-16}
Component & Parameter & L1  & L2 & L3 & L4& & L1  & L2 & L3 & L4&& L1  & L2 & L3 & L4  \\ 
 \hline
{\tt TBabs}   & $N_{\rm{H}}$ [10$^{21}$\,cm$^{-2}$]   &   \multicolumn{4}{c}{1.4 (F)} && \multicolumn{4}{c}{1.4 (F)} && \multicolumn{4}{c}{1.4 (F)}\\
\hline
{\tt diskbb}$_1$  & $kT_{\rm{in 1}}$\,[keV]          &\multicolumn{4}{c}{0.24$^{+0.01}_{-0.01}$} && \multicolumn{4}{c}{0.23$^{+0.01}_{-0.01}$} && \multicolumn{4}{c}{0.21$^{+0.01}_{-0.01}$} \\
 & $N_{1}$ [10$^{3}$]  & 15.8$^{+2.4}_{-1.9}$  & 8.1$^{+0.5}_{-0.5}$ & 3.2$^{+0.5}_{-0.5}$ & 0$^{+0.1}$  && 19.0$^{+3.3}_{-2.6}$  & 13.7$^{+2.1}_{-1.8}$ & 0$^{+0.4}$ & 0$^{+0.4}$ && 23.7$^{+4.7}_{-3.7}$  & 19.6$^{+4.0}_{-3.6}$ & 0$^{+0.8}$ & 0$^{+0.8}$\\
\hline
{\tt ThComp}  & $\Gamma$            & \multicolumn{4}{c}{1.56$^{+0.03}_{-0.03}$} && 1.64$^{+0.04}_{-0.04}$ & 1.63$^{+0.05}_{-0.05}$ & 1.91$^{+0.06}_{-0.05}$ & 1.32$^{+0.06}_{-0.06}$ && 1.67$^{+0.04}_{-0.04}$ & 1.64$^{+0.05}_{-0.05}$ & 2.20$^{+0.18}_{-0.13}$ & 1.48$^{+0.27}_{-0.13}$\\
& $kT_{\rm{e}}$\,[keV]     & \multicolumn{4}{c}{50 (F)} && \multicolumn{4}{c}{50 (F)} && \multicolumn{4}{c}{50 (F)} \\
{\tt diskbb}$_2$  & $kT_{\rm{in 2}}$\,[keV]      &\multicolumn{4}{c}{= $kT_{\rm{in 1}}$} && \multicolumn{4}{c}{= $kT_{\rm{in 1}}$} &&  = $kT_{\rm{in 1}}$ & 0.34$^{+0.11}_{-0.09}$ & 0.32$^{+0.03}_{-0.03}$ & 0.83$^{+0.39}_{-0.29}$ \\
 & $N_{2}$ [10$^{3}$]  & 8.1 $^{+1.7}_{-1.3}$ &  6.1$^{+1.2}_{-0.9}$ & 4.9$^{+1.0}_{-0.8}$  & 5.2$^{+0.9}_{-0.7}$    && 12.5$^{+2.8}_{-2.2}$ & 8.9$^{+2.1}_{-1.6}$ & 9.7$^{+1.6}_{-1.3}$ & 5.7$^{+0.9}_{-0.7}$ && 16.8$^{+3.0}_{-3.3}$     & 2.1$^{+3.4}_{-1.4}$ & 2.8$^{+1.1}_{-0.8}$ & 0.07$^{+0.30}_{-0.05}$ \\ 
\hline
\multicolumn{2}{c}{$\chi_\nu^{2}$} &  \multicolumn{4}{c}{318.8/160} && \multicolumn{4}{c}{232.7/157} && \multicolumn{4}{c}{185.9/154}\\
\hline
\end{tabular}
\label{tab:onec_model}
\end{table*}

\setlength{\tabcolsep}{8.1pt}
\begin{table*}
\centering
\caption{The joint fit results for model D with disc blackbody and two Comptonization zones for each of the \emph{rms} components and the time-averaged spectrum, S$_{\rm av}$. The {\sc Xspec} notation is \texttt{TBabs(diskbb$_1$+ThComp$_1$(diskbb$_2$)+ThComp$_2$(ThComp$_1$(diskbb$_3$))+Gaussian)}. The single {\tt diskbb} emission is split between three parts with the same $kT_{\rm in}$ and different normalizations, $N_1$, $N_2$ and $N_3$. The Roman numbers correspond to the zones shown in Fig.\ \ref{geo}. The component fluxes, $F$, are unabsorbed and bolometric (measured in the energy range 0.001-2000keV).
}
\begin{tabular}{ c c c c c c c c  }
 \hline
 & &    & \multicolumn{5}{c}{D}   \\
  \cline{4-8}    
&Component & Parameter & L1  & L2 & L3 & L4& S$_{\rm{av}}$  \\ 
 \hline
&{\tt TBabs}   & $N_{\rm{H}}$ [10$^{21}$\,cm$^{-2}$]   & \multicolumn{5}{c}{1.4$^{+0.1}_{-0.1}$} \\
\hline
I &{\tt diskbb}$_1$  & $kT_{\rm{in}}$\,[keV]        & \multicolumn{5}{c}{0.18$^{+0.01}_{-0.01}$} \\
    &    & $N_{1}$ [10$^{3}$]  & 52.7$^{+10.2}_{-9.8}$  & 34.8$^{+9.8}_{-6.4}$ & 0$^{+3.2}$ & 0$^{+1.4}$ & 100.6$^{+30.1}_{-18.8}$   \\
    &\multicolumn{2}{c}{$F\,[10^{-9}\,{\rm erg\,cm}^{-2}\,{\rm s}^{-1}$]}  & 1.21$^{+0.23}_{-0.22}$  & 0.79$^{+0.23}_{-0.15}$ & 0$^{+0.07}$ & 0$^{+0.03}$ & 2.31$^{+0.69}_{-0.43}$  \\    
\hline
II &{\tt ThComp}$_1$  & $\Gamma_{1}$            & \multicolumn{5}{c}{1.65$^{+0.14}_{-0.13}$} \\
   &     & $kT_{\rm{e 1}}$\,[keV]     & \multicolumn{5}{c}{0.34$^{+0.04}_{-0.03}$}    \\
 & {\tt diskbb}$_2$ & $N_{2}$ [10$^{3}$]  & 8.9$^{+3.8}_{-2.2}$  & 7.1$^{+2.9}_{-1.7}$ & 10.6$^{+1.8}_{-1.5}$ & 0$^{+0.5}$ & 8.9$^{+4.2}_{-2.4}$  \\
      &\multicolumn{2}{c}{$F\,[10^{-9}\,{\rm erg\,cm}^{-2}\,{\rm s}^{-1}$]}  & 0.47$^{+0.21}_{-0.11}$  & 0.38$^{+0.19}_{-0.11}$ & 0.57$^{+0.12}_{-0.09}$ & 0$^{+0.03}$ & 0.48$^{+0.27}_{-0.16}$  \\  
\hline
III &{\tt ThComp}$_2$  & $\Gamma_2$            & \multicolumn{5}{c}{1.49$^{+0.01}_{-0.01}$} \\
   &     & $kT_{\rm{e2}}$\,[keV]     & \multicolumn{5}{c}{50 (F)} \\
&{\tt diskbb}$_3$ & $N_{3}$ [10$^{3}$]  & 13.5$^{+2.9}_{-1.8}$ & 9.9$^{+2.1}_{-1.3}$ & 4.7$^{+1.3}_{-0.8}$ & 9.3$^{+1.3}_{-1.1}$ & 48.6$^{+9.8}_{-6.1}$  \\ 
      &\multicolumn{2}{c}{$F\,[10^{-9}\,{\rm erg\,cm}^{-2}\,{\rm s}^{-1}$]}  & 14.61$^{+3.16}_{-1.93}$  & 10.75$^{+8.48}_{-2.06}$ & 5.08$^{+1.28}_{-0.77}$ & 10.06$^{+1.23}_{-1.05}$ & 52.64$^{+9.56}_{-5.91}$  \\
\hline
& {\tt Gaussian}  & $E$\,[keV]  &\multicolumn{5}{c}{6.5$^{+0.1}_{-0.1}$}  \\
 &   & $\sigma$\,[keV] & \multicolumn{5}{c}{0.39$^{+0.04}_{-0.08}$}  \\
  &  & $N\,[10^{-3}\,{\rm cm}^{-2}\,{\rm s}^{-1}$] & 0$^{+1.06}$ & 0$^{+4.3}$ & 0$^{+4.3}$ & 0$^{+4.3}$ & 4.3$^{+0.8}_{-0.8}$   \\
\hline
& \multicolumn{2}{c}{$\chi_\nu^{2}$} &  \multicolumn{4}{c}{157.793/150} & 213.571/255 \\
& \multicolumn{2}{c}{} &  \multicolumn{5}{c}{370.4/415} \\
\hline
\end{tabular}
\label{tab:twoc_model}
\end{table*}

\section{Discussion}
\label{sec:discussion}

In our approach, spectral components derived from variability on different time scales are thought to originate from regions of the accretion flow related to these time scales. Generally, we expect the characteristic variability time scale to decrease with the decreasing radial distance from the BH. However, connecting variability time scales to radii is far from simple. Detailed spectro-timing models for Cyg X-1 and GX 339--4, in which such radii were identified, were performed by \citet{Mahmoud2018MNRAS.480.4040M} and \citet{Mahmoud2019MNRAS.486.2137M}. In particular, the former work found evidence for the existence of discrete regions of enhanced turbulence corresponding to humps in the power spectra. 

Here, we simply compare the characteristic Keplerian frequency, $\nu_{\rm K}\equiv (GM)^{1/2}/(2\upi R^{3/2})$ to the characteristic frequencies of our power spectrum, to identify the corresponding characteristic radii. Since this frequency corresponds to the fastest possible variability, it sets an upper limit on the actual emitting radii. This can be written as
\begin{equation}
\frac{R}{R_{\rm g}}\lesssim \frac{c^2}{\left(2\upi G M\nu\right)^{2/3}}\approx 250 \left(\frac{M}{8\msun}\right)^{-2/3} \left(\frac{\nu}{1\,{\rm Hz}}\right)^{-2/3}.
\label{RK}
\end{equation}
For the highest peak frequency in the power spectrum, $\approx$2.5\,Hz (L4 in Table \ref{tab:lore}), we have $R/R_{\rm g}\lesssim 140$. We can compare it with the radius based on the disc blackbody normalization, using its definition in {\sc Xspec} (stress free inner boundary condition is not included as defined in {\tt diskbb}).We used the total normalization in the time-averaged spectrum in model D, ($N_1+N_2+N_3$). However, this might overestimate the real amount of seed photons from the disc if other sources of seed photons are present \citep[e.g. with little overlap between the disc and the hot flow synchrotron emission might represent this additional source, e.g.][]{Poutanen2018A&A...614A..79P, Veledina2011ApJ...737L..17V}. Therefore, our fits allow us to put an upper limit on $R_{\rm in}$, that corresponds to the real value if the thermal photons from the disc are the only source of seed photons. We find,

\begin{align}
R_{\rm in}&\lesssim 5.3^{+0.5}_{-0.3} \left(\frac{\kappa}{1.7}\right)^2 \frac{d}{3\,{\rm kpc}} \left(\frac{\cos i}{\cos 65\degr}\right)^{-1/2} 10^7\,{\rm cm}\approx \\ 
&\approx 45^{+4}_{-3} \left(\frac{M}{8\msun}\right)^{-1} \left(\frac{\kappa}{1.7}\right)^2 \frac{d}{3\,{\rm kpc}} \left(\frac{\cos i}{\cos 65\degr}\right)^{-1/2} R_{\rm g},
\label{rin_rg}
\end{align}
where $\kappa\approx 1.5$--1.9 is the disc blackbody colour correction (e.g. \citealt{Davis05}). This satisfies the constraint (\ref{RK}), and shows the disc may be significantly truncated. Then, the hot accretion flow zones II and III can fit downstream of $R_{\rm in}$, showing the self-consistency of this aspect of model D.

Our findings are valid under the assumption that that variability is only due to changes in the normalization of each spectral component. In reality the situation might be much more complex, e.g. as a consequence of spectral pivoting associated with each Comptonization zone. A more rigorous approach implies building appropriate models of the time-variable spectrum, e.g. by perturbing the parameters of the time-averaged spectrum \citep{Gierlinski_Zdziarski2005MNRAS.363.1349G} or by solving the Kompaneets equation \citep{Karpouzas2020MNRAS.492.1399K, GarciaF2021MNRAS.501.3173G}. Nonetheless, we checked whether spectral pivoting may have a strong influence on our results, by calculating the fractional \emph{rms} spectra of of each Lorentzian (Fig.~\ref{fig:erms} - left panel). We observe that the fractional rms generally flattens out at energies $\gtrsim$3 keV. This suggests that high energy spectral components predominantly vary in normalization only, thus supporting our initial assumption. These caveats apply also when modeling disk blackbody emission. Indeed, the changes of the blackbody flux are connected to the changes of its temperature, so assuming only variations of normalization results in overestimating the real temperature in fits of \emph{rms} spectra \citep{vanParadijs1986A&A...157L..10V}.

%It seems that the L4 might be explained in terms of variations of the normalization of the inner Comptonization component only. But the fractional \emph{rms} spectra of lowest frequency Lorentzians (L1 and L2) are more complicated, and that may be caused by the outer Comptonization varying in normalization and/or slope plus inner Comptonization varying in normalization, or the inner Comptonization varying in slope (with a pivot point at high energies). On the other hand L3 looks the most complex (with not variable low energy part, and rise of variability at the highest energies) and is hard to explain. We conclude that our assumption is correct, especially at high energies. 

The question our research directly addressed was the structure of the accretion flow. Detailed modelling of Comptonization in the accretion flow appears not unique. Clearly, this flow does not consist of a single uniform plasma cloud, as shown by the failure of our model A to describe the \emph{rms} spectra. Our model B (Table \ref{tab:onec_model}) shows that a significant improvement is achieved when allowing the slopes of the \emph{rms} spectra to differ between the components, in agreement with what is seen in Fig.\ \ref{fig:erms}. This can be due to the parameters of the Comptonizing cloud, namely the optical depth and electron temperature, evolving along the accretion flow, and forming distinct zones. 

Our model C points to the average energy of the seed photons for Comptonization to vary between the \emph{rms} spectra of each Lorentzian component. The inner temperature for L1 is weakly constrained, but consistent with being the same as that of the directly observed disc blackbody; those for L2 and L3 are somewhat larger, and the seed photons for L4 have significantly higher temperature, $kT_{\rm in}\approx 0.5$--1.2\,keV. This may suggest that photons upscattered in one zone serve as the seed photons for the subsequent zones (as in the models of \citealt{Axelsson2018MNRAS.480..751A} and \citealt{Mahmoud2018MNRAS.480.4040M}). Alternatively, there could be cold clumps formed by some instabilities in the inner parts of the hot flow (e.g. \citealt{Wu16}; see \citealt{Poutanen2018A&A...614A..79P} for consequences for radiative cooling). Still, even this model leaves significant residuals after the fit, with the actual spectrum of L4 having a sharper low-energy cutoff than that in the fitted model, see Fig.\ \ref{fig:resi}. Disc blackbody emission is significant in the \emph{rms} spectra of the two lower frequency Lorentzians (L1 and L2), meaning that the disc strongly varies on long time scales ($\nu_{max}\lesssim$1\,Hz, Table \ref{tab:onec_model} model B and C). On the other hand, the direct disc blackbody emission is not observed in the \emph{rms} spectra of L3 and L4, meaning that the disc does not contribute significantly to the high frequency variability components ($\nu_{max}>$1Hz). At the same time, Comptonization emission varies on all time scales, but its fastest-varying component, L4, is also the hardest, with no detectable variability at energies $\lesssim$1\,keV (Fig.\ \ref{fig:erms}, Table \ref{tab:onec_model}). However, the spectral evolution over Lorentzians with increasing peak frequency is relatively complex, with L1 and L2 having very similar shapes and L3 being the softest in the $\approx$1--2\,keV range and then hardening. This appears to be a more complex behaviour than that found for Cyg X-1 \citep{Axelsson2018MNRAS.480..751A,Mahmoud2018MNRAS.480.4040M}. However, those papers based their analysis on the data at photon energies at $\geq$3\,keV, where indeed our \emph{rms} spectra show more uniformity (Fig.\ \ref{fig:erms}). 

In model D, we explored an alternative scenario in which each of the \emph{rms} spectra is fitted by two Comptonization zones. Following results from fits of model B and C, we ascribe the disc blackbody emission to the outer disc zone I (Fig.~\ref{geo}). The electrons in the outer Comptonization zone (zone II in Fig.~\ref{geo}) upscatter the disc blackbody photons, while those in the inner Comptonization zone (zone III in Fig.~\ref{geo}) upscatter the photons created in the outer Comptonization zone. This follows the hints for such complexity from the results of the fitting of the previous models B and C.

Model D can simultaneously describe the time-averaged and \emph{rms} spectra quite well. The fractional variability of the directly observed disc blackbody emission is as high as $\gtrsim$60 per cent (model D; Table \ref{tab:lore}). This value is consistent with estimates for GX 339-4 in the hard state \citep{De_Marco_2015}. The variability of the disc blackbody photons is almost entirely in the low-frequency components L1 and L2. Still, the components L1 and L2 also show variability in the Comptonization zones. The spectrum L3 has significant contributions from both the outer and inner Comptonization zones but none from the disc, while L4 is formed almost exclusively in the inner zone.

The behaviour described above strongly supports the picture of propagating fluctuations, with the dominant variability moving from the outer disc through the outer hot flow to the inner one while the dominant characteristic variability frequency increases. Our results agree with the detection of hard X-ray lags in this source \citep{Kara2019Natur.565..198K, DeMarco21}. Indeed, such lags are commonly ascribed to propagation of mass accretion rate fluctuations in the accretion flow \citep{Lyubarskii1997MNRAS.292..679L}. In order to produce lags, the perturbations have to propagate through a spectrally inhomogeneous region, with the hardest spectrum produced at the smallest radii \citep{Kotov2001MNRAS.327..799K}, as we find from our fits.

Still, this model has a number of caveats. It shows significant residuals at the low-energy cutoff of the spectrum L4, similarly as in the previous models. Such residuals suggest that the emission from zone III may be underestimated in the spectrum of L1, L2, and L3. This would ultimately lead to overestimate emission from zone II, resulting in this component exceeding the corresponding time average contribution. A solution for this would be to include further stratification of the hot flow, but our data are statistically not sufficient for such complex models. Moreover, since we assumed that there are only two Comptonization zones, the \emph{rms} spectrum of each Lorentzian is modelled assuming the same slope for each Comptonization zone, and only the relative normalizations vary. While this provides a good fit to the data at $\leq$10\,keV, it fails to describe the hardening observed at $\gtrsim$10\,keV in this source \citep{Zdziarski21b}. Such hardenings are also observed in other BH XRBs, e.g. Cyg X-1 \citep{Nowak2011ApJ...728...13N} or XTE J1752--223 \citep{Zdziarski21}. We note that, if the variations of spectral shape are only in normalization then the outer Comptonization zone has a very low electron temperature, $kT_{\rm e}\approx 0.34^{+0.04}_{-0.03}$\,keV. This means that this is likely to be a warm corona above the disc rather than a hot flow. The implied Thomson optical depth is $\tau_{\rm T}\approx 48$ (as follows for the used spectral model from eq.\ 14 of \citealt{Z20_thcomp}), which is quite unrealistic. Looking at the spectral components in Fig.\ \ref{geo}, we see that this component resembles quite well another blackbody component rather than a genuine Comptonization zone. While this points to a deficiency of our model, more model complexity could not be reasonably constrained by our data. Similar spectral decomposition was found for another X-ray binary Cyg X-1 \citep{DiSalvo2001ApJ...547.1024D, Basak2017MNRAS.472.4220B, Frontera2001ApJ...546.1027F, Makishima2008PASJ...60..585M, Nowak2011ApJ...728...13N, Yamada2013PASJ...65...80Y}.

\section{Conclusions}
We studied the spectral structure of the Comptonization region in the hard state of MAXI J1820+070 using \nicer data. 
To this aim we extracted the energy spectra of each Lorentzian describing the humped shape of the PSD of the source. The distribution over timescales of these variability components is thought to resemble the spatial distribution of energy dissipation zones in the hot flow, with the highest frequency Lorentzians corresponding to variability produced in the innermost regions. 

The main result of our analysis is the evidence of a spectral stratification of the hot flow, which can be clearly appreciated in a model-independent way by simply looking at the significant changes in the \emph{rms} spectra of the different Lorentzians (Fig.~\ref{fig:erms}). Going from the lowest to the highest frequency Lorentzian (L1 and L4 respectively) a net hardening of the hard X-ray spectrum is observed.

We modeled the \emph{rms} spectra of each Lorentzian in order to characterize the spectral structure of the source. Our simple models B and C (Sect. \ref{sec:fits} and Table \ref{tab:onec_model}) show that the slope of the Comptonization component significantly changes among Lorentzians. This means that the physical properties of the Comptonization region change as a function of radius. In particular, we found that the highest frequency Lorentzian L4 has a significantly harder spectrum than L3 (Table \ref{tab:onec_model}). The two lowest frequency Lorentzians L1 and L2 do not show significant changes of spectral index and have a softer/harder spectrum than L4/L3 (Table \ref{tab:onec_model}). Also, our model C provides evidence for hotter seed photons ($kT_{\rm in}=0.85^{+0.41}_{-0.29}$ keV) in the fastest variability component L4 compared to the other components ($kT_{\rm in}\sim0.3$ keV).

Consistent with the above findings, we then describe the hot flow in a self consistent way by two Comptonization zones and applied this model simultaneously to the \emph{rms} spectra of each Lorentzian and the time-averaged spectrum of the source. The model (D, Table \ref{tab:twoc_model}) comprises an outer Comptonization region fueled by thermal photons from the cool disc, and an inner Comptonization region fueled by a fraction of the upscattered photons from the outer Comptonization region. We find that this model can describe data spectra well.

\section*{Acknowledgements}
The authors acknowledge Mariano Mendez and the referee, Chris Done, for helpful discussions and suggestions which significantly improved the paper. We have benefited from discussions during Team Meetings of the International Space Science Institute in Bern, whose support we acknowledge. We also acknowledge support from the Polish National Science Centre under the grants 2015/18/A/ST9/00746, 2019/35/B/ST9/03944, the European Union's Horizon 2020 research and innovation programme under the Marie Sk{\l}odowska-Curie grant agreement No.\ 798726, and from Ram{\' o}n y Cajal Fellowship RYC2018-025950-I. 

\section*{Data availability}
The data underlying this article are available in HEASARC, at \url{https://heasarc.gsfc.nasa.gov/docs/archive.html}. 
\bibliographystyle{mnras}
\bibliography{1820_rms} 

\bsp	% typesetting comment
\label{lastpage}
\end{document}